\documentstyle[emlines2]{article}
\title{
       Symmetry for Nearest Stars and N-body Problem with
       Delay for Interaction.}
\author{
\\
      R.T.Faizullin \\ \\
     Omsk State University, Pr. Mira 55a, 644077, Omsk-77, Russia\\ \\
 \\}
\date{}
\begin{document}
\maketitle

 In natural way one can distinguish regular structures of the visible
stars
 as constellations. But there are also \it{ natural} \rm coordinate
 systems which can give us information on more subtle structure and
 associations between the functions of coordinates, moments and
masses.
 It is also possible to introduce an order, or, in other words,
 enumeration of the stars, which is associated with these functions.
 These are ($\alpha, \delta$), ($l,b$), ($\lambda, \beta$)
 moment related coordinate systems, and enumeration according to the
visual
 brightness, or, more properly, according to the physical
characteristic
 V $[1]$.  For the first 20 stars their
 numbers and ($\alpha, \delta$) coordinates are given below:

  1 $\alpha Cma $           6h 43   -16 -35;
  2 $\alpha Car $          6h 23   -52 -40;
  3 $\alpha Bool$            14h 13   19 27;
  4 $\alpha Lyr$                    18h 35   38 44;
  5 $\alpha Cen $            14h 36  -60 -38;
  6 $\alpha Aur$             5h 13    45 57;
  7 $\beta Ori $           5h 12    -8 -15;
  8 $\alpha Cmi$                    7h 37     5 21;
  9 $\alpha Ori$                    5h 52     7 24;
  10 $\alpha Eri$                   1h 36   -57 -29;
  11$\beta Cen$              14h 0   -60 -8;
  12 $\alpha Aql$            19h 48    8 44;
  13 $\alpha Cru$                   12h 24  -62 -49;
  14 $\alpha Tau$                   4h 33    16 25;
  15$\alpha Sco$                    16h 26  -26 -19;
  16 $\alpha Vir$                   13h 23  -10 -54;
  17 $\beta Gem$            7h 42    28 9;
  18 $\alpha Psa$                   22h 55  -29 -53;
  19  $\beta Cru$                   12h 45  -59 9;
  20 $\alpha Cyg$             20h 40   45 6;

 As an example, let us consider the stars with numbers
 5, 6, 8, 12, 14, 15 on ($\lambda, \beta$) coordinate
 plane -- Fig.1. It is an example of graph of a new kind:
 \it{ each point i belongs to a straight line drawn through
 other two points j,k from the list, the sum of j and k being
 equal to 20}\rm.

 Note that these are really \it{the nearest} \rm extremally massive
objects
 (may be, with exception of 15), and, besides that, the
 stars 5, 8, 12, 6, 15, 14 also provide a regular structure
 on (l,b) plane -- Fig.2.

 Using the coordinates of the nearest stars, we can construct another
 regular structure, or, in other words, give an example of graph of a
new
 kind, using the representation of the stars with numbers
 5, 8, 9, 12, 13, 17 in the galactic coordinate system -- Fig.3.
 For these numbers and their complements with respect to 20 we can
 conclude that:
       \it  for each numbered node i there is an interval whose length
 is proportional to the number i;
        there are also intervals whose lengths are proportional
 to the complements with respect to 20;
        each such interval is connected with the corresponding
 node (the plot is extended twice along the axis b). \rm

 When moving period by period along the line (10,3) in the plane
 ($\lambda, \beta$), the coordinates of the stars with numbers
 25, 15, 6, 5, 11, 20, 2 concentrate closely to this line. And what
is more,
 there is "almost divisibility by five" --
 the stars with numbers 5, 10, 15, 20, 25, 30, 35, 40, 45, 50
 (and 6, 11, 19, 21, 29, 31, 36, 39) concentrate only to certain
lines
 ( more then two points on one line in all the coordinate systems).

 It is well known that there is a structure of new $O,B$ and $AO$
 stars; Herschel (1847) was the first to note that, and statistically
 convincing arguments were given by Guld (1879), that is so called
Guld zone.
 Some visual phenomena associated with the enumeration according to
the
 brightness may be indicative of existence of a more subtle structure.

 With rare exception, most part of $O,B$ and $AO$ stars (brighter
than $6.0^m$)
 is concentrated near the straight lines drawn
 through pairs of the stars with numbers 2, 3,
 5, 6, 7, 10, 11, 15; that is true
 for ($l,b$), ($\alpha,\delta$), ($\lambda,\beta$) coordinate planes.
 This fact is illustrated on Figs.4,5.

 This property is true also for the group 2, 6, 7, 10, 15 for earlier
times
 ($-10^6,-2*10^6$ years) on ($l,b$) plane.

 Perhaps this phenomenon is not completely visual, there is
 a substructure of stars of $\beta Per$ (EA) and $\beta Lyr$ (EB)
types,
 which concentrate only to $some$ of the lines in all the coordinate
systems.
 It is an independent test, in the average the brightness of these
stars is
 $8-9^m$. $K,M$ giants also gravitate towards these lines, but their
positions
 are more smeared than those of $O,B$ stars.

 When considering the group of the stars with numbers
 3, 4, 6, 7, 8, 9, 10, 11, 13, 17,
 we also can see that the most part of the bright galactic objects
 (brighter than $6^m,11^m$) concentrate to analogous system of
straight
 lines.

 This can be a result of the view from the inside of
 interacting vortices, where the points 7, 5, 10, 15 lie near the
singular
 points of the vortices.

  What's kind of invariance can be choosen for description of these
  phenomenon? If we consider N-body problem with delay
  for gravitational interaction then we have some limitation
  for the distances and velosities.
   It's follows from deduction of the energy integral
   for Newton law with delay.
    Velosity should be less then signal velosity $c$ for interactioned
   bodies and distances greater  then gravitational radius in every
time.
    Moreover, there can be upper limitation for distances
    and velosities for bodies are greater then velosities for standard
    model.

     Computer simulation for hierachical coupling $[2]$ with
     delay  for gravitational interactions may be give to us
     same configurations as "eigenfunctions" for the arbitrary
      initial density distribution.

{\ }

{\bf References}

1. P.G.Kulikovski, {\it Stars astronomy}, Nauka ,Moskow(1985)

2. R.T.Faizullin, {\it Massexchange in globula cloud}, Omsk St.Univ.,

  Omsk(1994)

\newpage
\unitlength=1.00mm
\special{em:linewidth 0.4pt}
\linethickness{0.4pt}
\begin{picture}(154.00,140.00)
\emline{140.00}{70.00}{1}{140.00}{140.00}{2}
\emline{140.00}{70.00}{3}{10.00}{70.00}{4}
\put(69.00,111.00){\makebox(0,0)[cc]{5}}
\
\put(68.08,109.00){\circle{1.00}}
\put(38.00,90.00){\makebox(0,0)[cc]{6}}
\
\put(36.27,88.64){\circle{1.00}}
\put(45.00,91.00){\makebox(0,0)[cc]{8}}
\
\put(43.15,89.66){\circle{1.00}}
\put(82.00,104.00){\makebox(0,0)[cc]{12}}
\
\put(80.42,102.36){\circle{1.00}}
\put(36.00,63.00){\makebox(0,0)[cc]{14}}
\
\put(33.77,60.74){\circle{1.00}}
\put(71.00,73.00){\makebox(0,0)[cc]{15}}
\
\put(69.81,70.81){\circle{1.00}}
\put(43.00,25.00){\makebox(0,0)[cc]{Fig.1 $\lambda, \beta$ plane,
puzzle with }}
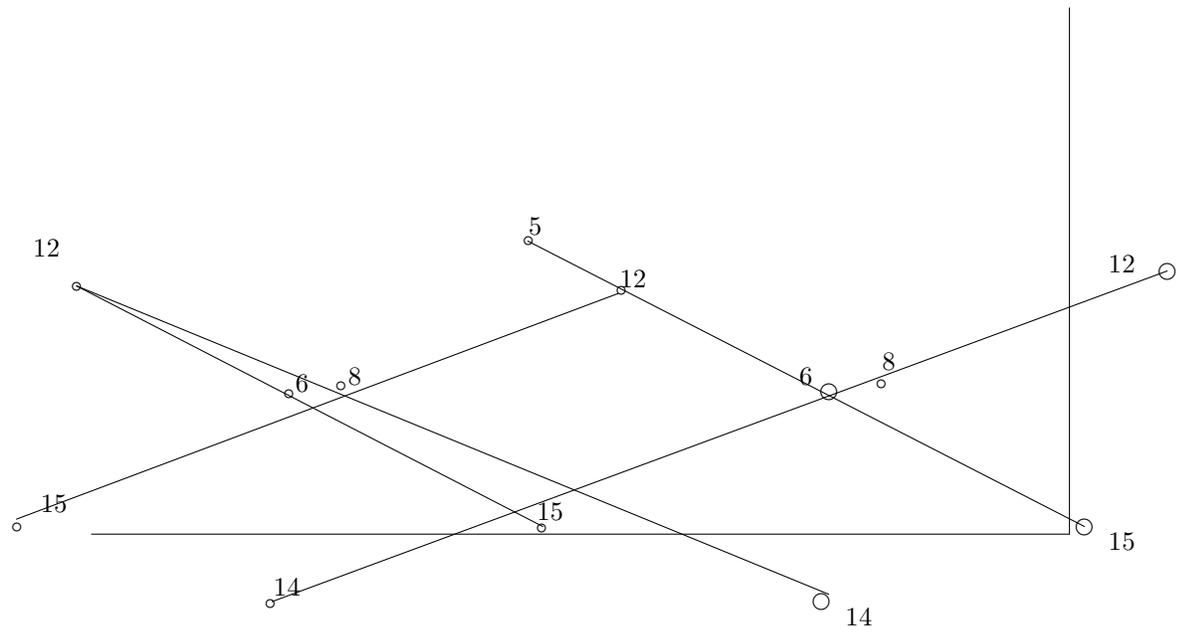
\put(43.00,20.00){\makebox(0,0)[cc]{numbers 5,6,8,12,14,15 and number
20}}
\
\put(8.00,103.00){\circle{1.00}}
\put(4.00,108.00){\makebox(0,0)[cc]{12}}
\put(0.10,71.00){\circle{1.00}}
\put(5.00,74.00){\makebox(0,0)[cc]{15}}
\put(115.00,90.00){\circle{1.00}}
\put(116.00,93.00){\makebox(0,0)[cc]{8}}
\emline{0.00}{72.00}{5}{80.00}{102.00}{6}
\emline{8.00}{103.00}{7}{70.00}{71.00}{8}
\put(153.00,105.00){\circle{2.00}}
\emline{34.00}{61.00}{9}{153.00}{105.00}{10}
\put(147.00,106.00){\makebox(0,0)[cc]{12}}
\put(142.00,71.00){\circle{2.00}}
\emline{68.00}{109.00}{11}{142.00}{71.00}{12}
\put(107.00,61.00){\circle{2.00}}
\emline{8.00}{103.00}{13}{108.00}{62.00}{14}
\put(112.00,59.00){\makebox(0,0)[cc]{14}}
\put(147.00,69.00){\makebox(0,0)[cc]{15}}
\put(108.00,89.00){\circle{2.00}}
\put(105.00,91.00){\makebox(0,0)[cc]{6}}
\end{picture}
\newpage
\unitlength=1.00mm
\special{em:linewidth 0.4pt}
\linethickness{0.4pt}
\begin{picture}(146.00,140.00)
\emline{20.00}{80.00}{1}{140.00}{80.00}{2}
\emline{140.00}{80.00}{3}{140.00}{140.00}{4}
\put(40.00,62.00){\circle{2.00}}
\put(93.00,82.00){\circle{2.00}}
\put(145.00,106.00){\circle{2.00}}
\emline{40.00}{62.00}{5}{145.00}{106.00}{6}
\put(112.00,62.00){\circle{2.00}}
\put(73.00,106.00){\circle{2.00}}
\emline{73.00}{106.00}{7}{112.00}{62.00}{8}
\put(73.00,109.00){\makebox(0,0)[cc]{8}}
\put(114.00,57.00){\makebox(0,0)[cc]{12}}
\put(91.00,76.00){\makebox(0,0)[cc]{5}}
\put(33.00,56.00){\makebox(0,0)[cc]{12}}
\put(145.00,110.00){\makebox(0,0)[cc]{8}}
\put(135.00,131.00){\makebox(0,0)[cc]{b}}
\put(23.00,73.00){\makebox(0,0)[cc]{l}}
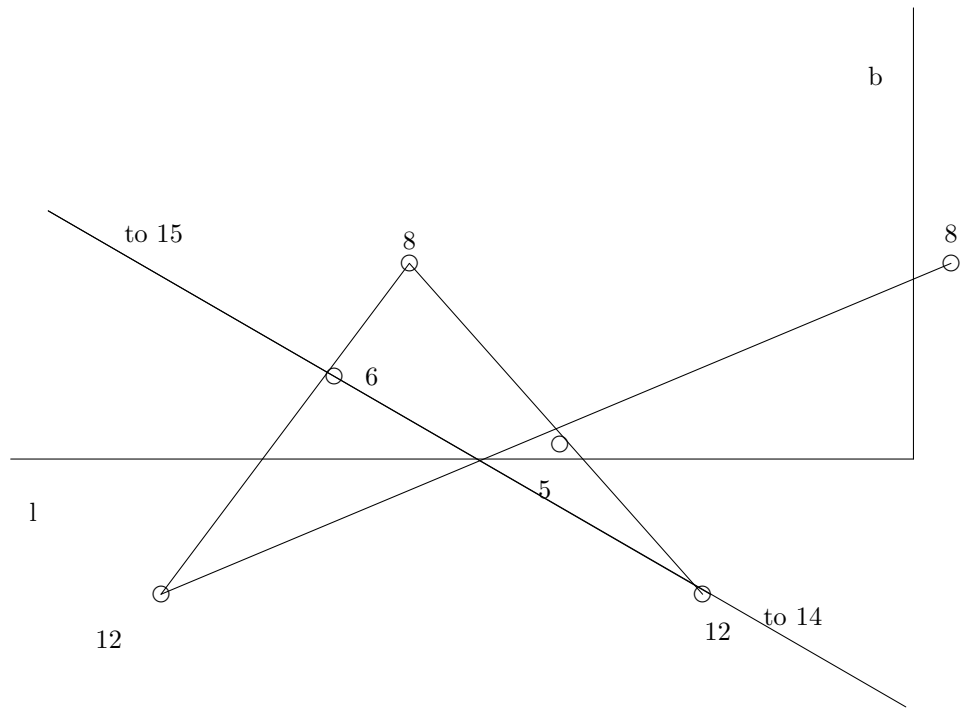
\put(47.00,40.00){\makebox(0,0)[cc]{Fig.2 Puzzle 12,5,8 on (l,b)
plane.}}
\emline{40.00}{62.00}{9}{73.00}{106.00}{10}
\put(63.00,91.00){\circle{2.00}}
\emline{113.00}{62.00}{11}{25.00}{113.00}{12}
\emline{25.00}{113.00}{13}{139.00}{47.00}{14}
\put(68.00,91.00){\makebox(0,0)[cc]{6}}
\put(124.00,59.00){\makebox(0,0)[cc]{to 14}}
\put(39.00,110.00){\makebox(0,0)[cc]{to 15}}
\end{picture}
\newpage
\unitlength=1.00mm
\special{em:linewidth 0.4pt}
\linethickness{0.4pt}
\begin{picture}(140.00,150.00)
\emline{10.00}{80.00}{1}{140.00}{80.00}{2}
\emline{140.00}{80.00}{3}{140.00}{150.00}{4}
\put(133.00,146.00){\makebox(0,0)[cc]{b}}
\put(17.00,93.00){\circle{2.00}}
\put(14.00,70.00){\circle{2.00}}
\put(6.00,103.00){\circle{2.00}}
\put(58.00,81.00){\circle{2.00}}
\put(56.00,80.00){\circle{2.00}}
\put(104.00,71.00){\circle{2.00}}
\emline{59.00}{81.00}{5}{56.00}{80.00}{6}
\emline{17.00}{93.00}{7}{105.00}{71.00}{8}
\emline{105.00}{71.00}{9}{6.00}{103.00}{10}
\emline{6.00}{103.00}{11}{17.00}{93.00}{12}
\emline{17.00}{93.00}{13}{56.00}{80.00}{14}
\emline{56.00}{80.00}{15}{58.00}{81.00}{16}
\emline{58.00}{81.00}{17}{17.00}{93.00}{18}
\emline{17.00}{93.00}{19}{14.00}{70.00}{20}
\emline{14.00}{70.00}{21}{6.00}{103.00}{22}
\emline{6.00}{103.00}{23}{56.00}{80.00}{24}
\emline{14.00}{71.00}{25}{29.00}{71.00}{26}
\emline{42.00}{71.00}{27}{55.00}{71.00}{28}
\emline{63.00}{71.00}{29}{79.00}{71.00}{30}
\emline{87.00}{71.00}{31}{103.00}{71.00}{32}
\emline{15.00}{71.00}{33}{58.00}{81.00}{34}
\put(109.00,68.00){\makebox(0,0)[cc]{11(12)}}
\put(61.00,76.00){\makebox(0,0)[cc]{3(5)}}
\put(52.00,75.00){\makebox(0,0)[cc]{13(13)}}
\put(8.00,106.00){\makebox(0,0)[cc]{15(17)}}
\put(20.00,87.00){\makebox(0,0)[cc]{8(8)}}
\put(13.00,63.00){\makebox(0,0)[cc]{12(9)}}
\put(125.00,75.00){\makebox(0,0)[cc]{l}}
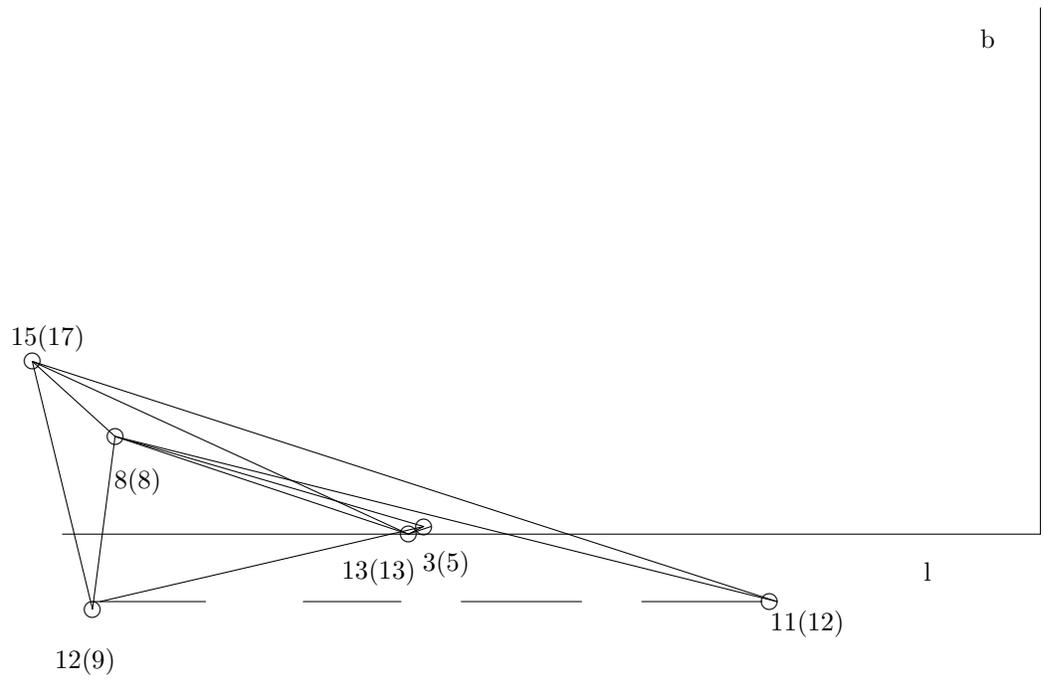
\put(70.00,48.00){\makebox(0,0)[cc]{Fig.3 Graph with associations
between enbumeration and distances.}}
\end{picture}
\newpage
\unitlength=1.00mm
\special{em:linewidth 0.4pt}
\linethickness{0.4pt}
\begin{picture}(144.05,142.00)
\emline{140.00}{70.00}{1}{140.00}{140.00}{2}
\put(63.00,75.00){\circle{0.30}}
\put(66.00,76.00){\makebox(0,0)[cc]{6}}
\put(82.00,45.00){\circle{0.30}}
\put(84.00,41.00){\makebox(0,0)[cc]{2}}
\put(72.00,45.00){\circle{0.30}}
\put(71.00,39.00){\makebox(0,0)[cc]{7}}
\put(100.00,85.00){\circle{0.30}}
\put(101.00,87.00){\makebox(0,0)[cc]{15}}
\emline{140.00}{70.00}{3}{10.00}{70.00}{4}
\put(93.00,71.00){\circle{0.30}}
\put(93.00,71.00){\circle{0.30}}
\put(93.00,73.00){\makebox(0,0)[cc]{5}}
\put(88.00,11.00){\circle{0.30}}
\put(89.00,14.00){\makebox(0,0)[cc]{10}}
\put(104.00,139.00){\circle{0.30}}
\put(107.00,139.00){\makebox(0,0)[cc]{3}}
\put(47.00,72.00){\circle{0.30}}
\put(48.00,74.00){\makebox(0,0)[cc]{20}}
\put(70.00,54.00){\circle{0.30}}
\put(71.00,56.00){\makebox(0,0)[cc]{25}}
\put(29.00,11.00){\circle{0.30}}
\put(28.00,14.00){\makebox(0,0)[cc]{10}}
\emline{29.00}{11.00}{5}{129.00}{115.00}{6}
\put(32.00,139.00){\circle{0.30}}
\put(34.00,142.00){\makebox(0,0)[cc]{3}}
\put(144.00,45.00){\circle{0.30}}
\put(144.00,49.00){\makebox(0,0)[cc]{7}}
\emline{100.00}{85.00}{7}{14.00}{63.00}{8}
\emline{32.00}{139.00}{9}{144.00}{46.00}{10}
\put(134.00,75.00){\circle{0.30}}
\put(135.00,78.00){\makebox(0,0)[cc]{6}}
\emline{135.00}{75.00}{11}{19.00}{107.00}{12}
\put(54.00,97.00){\circle{0.30}}
\put(54.00,99.00){\makebox(0,0)[cc]{50}}
\emline{88.00}{11.00}{13}{54.00}{97.00}{14}
\put(93.00,121.00){\circle{0.30}}
\put(95.00,123.00){\makebox(0,0)[cc]{16}}
\emline{63.00}{75.00}{15}{104.00}{139.00}{16}
\put(89.00,73.00){\circle{0.30}}
\put(87.00,75.00){\makebox(0,0)[cc]{19}}
\put(99.00,18.00){\circle{0.30}}
\put(102.00,14.00){\makebox(0,0)[cc]{30}}
\emline{144.00}{45.00}{17}{77.00}{4.00}{18}
\put(85.00,59.00){\makebox(0,0)[cc]{35}}
\put(83.00,62.00){\circle{0.30}}
\emline{82.00}{45.00}{19}{104.00}{140.00}{20}
\emline{29.00}{11.00}{21}{129.00}{104.00}{22}
\put(99.00,64.00){\circle{0.30}}
\put(101.00,60.00){\makebox(0,0)[cc]{40}}
\emline{104.00}{140.00}{23}{95.00}{15.00}{24}
\emline{88.00}{11.00}{25}{104.00}{139.00}{26}
\emline{72.00}{45.00}{27}{134.00}{75.00}{28}
\put(93.00,55.00){\circle{0.30}}
\put(95.00,52.00){\makebox(0,0)[cc]{43}}
\put(78.00,101.00){\circle{0.30}}
\put(78.00,105.00){\makebox(0,0)[cc]{47}}
\put(59.00,118.00){\circle{0.30}}
\put(62.00,120.00){\makebox(0,0)[cc]{33}}
\emline{32.00}{139.00}{29}{131.00}{29.00}{30}
\put(104.00,60.00){\circle{0.30}}
\put(107.00,59.00){\makebox(0,0)[cc]{36}}
\put(128.00,31.00){\circle{0.30}}
\put(127.00,26.00){\makebox(0,0)[cc]{49}}
\put(69.00,93.00){\circle{0.30}}
\put(68.00,95.00){\makebox(0,0)[cc]{17}}
\put(44.00,71.00){\circle{0.30}}
\put(93.00,93.00){\circle{0.30}}
\put(94.00,87.00){\circle{0.30}}
\put(89.00,88.00){\circle{0.30}}
\emline{82.00}{45.00}{31}{40.00}{109.00}{32}
\emline{40.00}{109.00}{33}{101.00}{16.00}{34}
\put(46.00,99.00){\circle{0.30}}
\put(74.00,57.00){\circle{0.30}}
\put(74.00,60.00){\makebox(0,0)[cc]{32}}
\put(123.00,37.00){\circle{0.30}}
\put(104.00,11.00){\circle{0.30}}
\put(108.00,11.00){\makebox(0,0)[cc]{18}}
\put(107.00,77.00){\circle{0.30}}
\put(102.00,61.00){\circle{0.30}}
\put(103.00,63.00){\makebox(0,0)[cc]{24}}
\put(104.00,84.00){\circle{0.30}}
\put(102.00,81.00){\circle{0.30}}
\put(67.00,66.00){\circle{0.30}}
\put(61.00,63.00){\makebox(0,0)[cc]{27}}
\put(96.00,73.00){\circle{0.30}}
\emline{84.00}{81.00}{35}{139.00}{94.00}{36}
\put(114.00,88.00){\circle{0.30}}
\put(116.00,86.00){\makebox(0,0)[cc]{4}}
\emline{74.00}{92.00}{37}{42.00}{43.00}{38}
\put(55.00,64.00){\circle{0.30}}
\emline{104.00}{139.00}{39}{61.00}{15.00}{40}
\put(65.00,27.00){\circle{0.30}}
\put(74.00,53.00){\circle{0.30}}
\put(77.00,61.00){\circle{0.30}}
\put(76.00,57.00){\circle{0.30}}
\put(100.00,71.00){\circle{0.30}}
\put(135.00,129.00){\makebox(0,0)[cc]{b}}
\put(16.00,72.00){\makebox(0,0)[cc]{l}}
\put(70.00,64.00){\circle{0.30}}
\put(99.00,71.00){\circle{0.30}}
\put(101.00,72.00){\makebox(0,0)[cc]{11}}
\put(67.00,141.00){\makebox(0,0)[cc]{Fig.4 Stars on (l,b) plane}}
\put(88.00,82.00){\circle{0.30}}
\put(41.00,22.00){\circle{0.30}}
\put(88.00,36.00){\circle{0.30}}
\put(72.00,50.00){\circle{0.30}}
\put(80.00,7.00){\circle{0.30}}
\put(95.00,87.00){\circle{0.30}}
\put(93.00,90.00){\circle{0.30}}
\put(78.00,99.00){\circle{0.30}}
\put(78.00,62.00){\circle{0.30}}
\end{picture}
\newpage
\unitlength=1.00mm
\special{em:linewidth 0.4pt}
\linethickness{0.4pt}
\begin{picture}(153.00,159.18)
\put(50.15,53.42){\circle{0.30}}
\put(49.15,17.33){\circle{1.50}}
\put(72.65,89.45){\circle{1.50}}
\put(85.75,108.73){\circle{0.30}}
\put(73.80,9.37){\circle{1.50}}
\put(45.65,115.95){\circle{1.50}}
\put(45.60,61.75){\circle{1.50}}
\put(52.85,75.35){\circle{0.30}}
\put(47.60,77.40){\circle{0.30}}
\put(34.80,12.52){\circle{1.50}}
\put(72.00,9.87){\circle{1.50}}
\put(89.40,78.73){\circle{0.30}}
\put(67.20,7.18){\circle{0.30}}
\put(43.65,86.42){\circle{0.30}}
\put(79.30,43.68){\circle{1.50}}
\put(70.15,59.10){\circle{0.30}}
\put(53.10,98.15){\circle{0.30}}
\put(98.75,40.12){\circle{0.30}}
\put(68.25,10.85){\circle{0.30}}
\put(92.00,115.10){\circle{0.30}}
\put(60.30,82.22){\circle{0.30}}
\put(50.85,41.10){\circle{0.30}}
\put(52.55,102.00){\circle{0.30}}
\put(82.50,32.93){\circle{0.30}}
\put(46.10,76.30){\circle{0.30}}
\put(67.40,13.17){\circle{0.30}}
\put(46.15,98.57){\circle{0.30}}
\put(57.65,0.48){\circle{0.30}}
\put(46.70,68.77){\circle{0.30}}
\put(96.25,22.80){\circle{0.30}}
\put(68.60,126.23){\circle{0.30}}
\put(46.90,68.03){\circle{0.30}}
\put(63.05,132.02){\circle{0.30}}
\put(40.05,119.68){\circle{0.30}}
\put(54.40,22.80){\circle{0.30}}
\put(85.05,35.58){\circle{0.30}}
\put(51.30,43.68){\circle{0.30}}
\put(71.30,119.57){\circle{0.30}}
\put(55.10,10.65){\circle{0.30}}
\put(82.70,27.03){\circle{0.30}}
\put(47.80,114.95){\circle{0.30}}
\put(49.75,86.45){\circle{0.30}}
\put(80.15,1.07){\circle{0.30}}
\put(56.15,15.48){\circle{0.30}}
\put(91.10,13.10){\circle{0.30}}
\put(49.00,52.07){\circle{0.30}}
\put(58.25,61.57){\circle{0.30}}
\put(36.85,66.80){\circle{0.30}}
\put(36.20,93.23){\circle{0.30}}
\put(35.45,159.03){\circle{0.30}}
\put(87.45,74.00){\circle{0.30}}
\put(30.35,98.95){\circle{0.30}}
\put(86.65,43.68){\circle{0.30}}
\put(65.35,84.70){\circle{0.30}}
\put(70.10,125.05){\circle{0.30}}
\put(74.50,144.25){\circle{0.30}}
\put(39.30,110.85){\circle{0.30}}
\put(36.10,112.20){\circle{0.30}}
\put(30.35,129.00){\circle{0.30}}
\put(31.95,126.00){\circle{0.30}}
\put(33.05,77.00){\circle{0.30}}
\put(68.00,22.00){\circle{0.30}}
\put(65.60,123.00){\circle{0.30}}
\put(72.20,34.00){\circle{0.30}}
\put(74.00,23.00){\circle{0.30}}
\put(73.65,28.00){\circle{0.30}}
\put(70.10,125.00){\circle{0.30}}
\put(74.40,97.00){\circle{0.30}}
\put(76.60,96.00){\circle{0.30}}
\put(78.00,48.00){\circle{0.30}}
\put(80.40,36.00){\circle{0.30}}
\put(83.00,31.00){\circle{0.30}}
\put(83.80,121.00){\circle{0.30}}
\put(91.00,110.00){\circle{0.30}}
\put(99.00,97.00){\circle{0.30}}
\put(93.85,132.47){\circle{0.30}}
\put(81.40,135.00){\circle{0.30}}
\put(82.65,82.57){\circle{0.30}}
\put(32.85,151.57){\circle{0.30}}
\put(34.40,123.75){\circle{0.30}}
\put(34.90,124.70){\circle{0.30}}
\put(35.85,97.63){\circle{0.30}}
\put(37.10,125.87){\circle{0.30}}
\put(38.10,117.93){\circle{0.30}}
\put(38.20,139.42){\circle{0.30}}
\put(39.20,110.75){\circle{0.30}}
\put(40.45,66.53){\circle{0.30}}
\put(42.00,97.98){\circle{0.30}}
\put(43.75,90.58){\circle{0.30}}
\put(44.05,59.23){\circle{0.30}}
\put(45.05,111.22){\circle{0.30}}
\put(45.30,109.52){\circle{0.30}}
\put(45.40,112.10){\circle{0.30}}
\put(46.30,105.33){\circle{0.30}}
\put(46.60,64.58){\circle{0.30}}
\put(46.70,97.23){\circle{0.30}}
\put(47.85,80.22){\circle{0.30}}
\put(49.95,50.38){\circle{0.30}}
\put(50.60,68.70){\circle{0.30}}
\put(56.05,89.20){\circle{0.30}}
\emline{140.00}{70.00}{1}{140.00}{140.00}{2}
\emline{140.00}{70.00}{3}{12.00}{70.00}{4}
\emline{49.00}{17.00}{5}{39.00}{149.00}{6}
\emline{73.00}{89.00}{7}{22.00}{140.00}{8}
\emline{22.00}{140.00}{9}{120.00}{42.00}{10}
\emline{72.00}{10.00}{11}{72.00}{150.00}{12}
\emline{34.00}{13.00}{13}{65.00}{148.00}{14}
\emline{80.00}{43.00}{15}{13.00}{78.00}{16}
\emline{35.00}{12.00}{17}{125.00}{50.00}{18}
\emline{46.00}{62.00}{19}{111.00}{133.00}{20}
\emline{73.00}{90.00}{21}{18.00}{142.00}{22}
\emline{49.00}{18.00}{23}{87.00}{5.00}{24}
\emline{73.00}{9.00}{25}{96.00}{137.00}{26}
\emline{34.00}{12.00}{27}{48.00}{140.00}{28}
\emline{72.00}{10.00}{29}{24.00}{100.00}{30}
\put(118.00,61.00){\circle{2.00}}
\put(152.00,44.00){\circle{2.00}}
\emline{152.00}{44.00}{31}{17.00}{123.00}{32}
\emline{35.00}{13.00}{33}{151.00}{44.00}{34}
\emline{118.00}{61.00}{35}{34.00}{13.00}{36}
\put(77.00,45.00){\circle{0.30}}
\put(77.00,44.00){\circle{0.30}}
\put(78.00,45.00){\circle{0.30}}
\emline{79.00}{43.00}{37}{64.00}{147.00}{38}
\emline{64.00}{147.00}{39}{84.00}{10.00}{40}
\put(78.00,48.00){\circle{2.00}}
\put(78.00,52.00){\circle{0.30}}
\put(77.00,53.00){\circle{0.30}}
\put(78.00,52.00){\circle{0.30}}
\put(74.00,28.00){\circle{0.30}}
\put(75.00,29.00){\circle{0.30}}
\put(77.00,29.00){\circle{0.30}}
\put(73.00,27.00){\circle{0.30}}
\put(72.00,26.00){\circle{0.30}}
\put(76.00,23.00){\circle{0.30}}
\put(77.00,24.00){\circle{0.30}}
\emline{34.00}{13.00}{41}{99.00}{144.00}{42}
\put(46.00,37.00){\circle{0.30}}
\put(48.00,35.00){\circle{0.30}}
\put(48.00,32.00){\circle{0.30}}
\put(45.00,34.00){\circle{0.30}}
\put(37.00,76.00){\circle{0.30}}
\put(38.00,74.00){\circle{0.30}}
\put(40.00,71.00){\circle{0.30}}
\put(35.00,78.00){\circle{0.30}}
\emline{46.00}{116.00}{43}{46.00}{0.00}{44}
\put(46.00,72.00){\circle{0.30}}
\put(45.00,76.00){\circle{0.30}}
\put(46.00,96.00){\circle{0.30}}
\put(46.00,92.00){\circle*{0.30}}
\put(78.00,101.00){\circle*{2.00}}
\put(87.00,118.00){\circle*{2.00}}
\put(97.00,138.00){\circle*{2.00}}
\put(100.00,144.00){\circle*{2.00}}
\put(46.00,112.00){\circle*{2.00}}
\put(45.00,114.00){\circle*{2.00}}
\put(46.00,106.00){\circle*{2.00}}
\put(47.00,100.00){\circle*{2.00}}
\put(68.00,86.00){\circle*{2.00}}
\put(68.00,86.00){\circle*{2.00}}
\put(82.00,101.00){\circle*{2.00}}
\put(90.00,109.00){\circle*{2.00}}
\put(89.00,109.00){\circle*{2.00}}
\put(89.00,109.00){\circle*{2.00}}
\put(89.00,97.00){\circle*{2.00}}
\put(89.00,96.00){\circle*{2.00}}
\put(87.00,90.00){\circle*{2.00}}
\put(89.00,93.00){\circle*{2.00}}
\put(88.00,88.00){\circle*{2.00}}
\put(88.00,88.00){\circle*{2.83}}
\emline{72.00}{10.00}{45}{102.00}{143.00}{46}
\put(93.00,101.00){\circle*{2.83}}
\put(96.00,112.00){\circle*{2.00}}
\put(44.00,119.00){\circle*{2.00}}
\put(39.00,123.00){\circle*{2.00}}
\put(33.00,130.00){\circle*{2.00}}
\put(25.00,137.00){\circle*{2.00}}
\put(19.00,142.00){\circle*{2.83}}
\put(103.00,125.00){\circle*{2.00}}
\put(108.00,130.00){\circle*{0.00}}
\put(108.00,131.00){\circle*{2.00}}
\put(72.00,99.00){\circle*{2.00}}
\put(70.00,107.00){\circle*{2.00}}
\put(109.00,56.00){\circle*{2.00}}
\put(96.00,76.00){\circle{2.00}}
\put(96.00,76.00){\circle*{2.00}}
\put(109.00,69.00){\circle*{2.00}}
\put(118.00,115.00){\circle{2.00}}
\emline{72.00}{90.00}{47}{146.00}{131.00}{48}
\put(112.00,112.00){\circle*{2.00}}
\put(116.00,113.00){\circle*{2.00}}
\put(93.00,104.00){\circle*{2.00}}
\put(42.00,111.00){\circle*{2.00}}
\put(41.00,130.00){\circle*{2.00}}
\put(39.00,132.00){\circle*{2.00}}
\put(43.00,91.00){\circle*{2.00}}
\put(44.00,81.00){\circle*{2.00}}
\put(47.00,69.00){\circle*{2.00}}
\put(92.00,127.00){\circle*{2.83}}
\put(96.00,134.00){\circle*{2.83}}
\put(35.00,23.00){\circle{0.30}}
\put(36.00,26.00){\circle{0.30}}
\put(77.00,34.00){\circle{0.30}}
\put(72.00,19.00){\circle{0.30}}
\put(72.00,24.00){\circle{0.30}}
\put(72.00,22.00){\circle{0.30}}
\put(65.00,13.00){\circle{0.30}}
\put(73.00,76.00){\circle{0.30}}
\put(62.00,29.00){\circle{0.30}}
\put(120.00,21.00){\makebox(0,0)[cc]{Fig.5 $\alpha, \beta$ plane}}
\put(106.00,10.00){\circle*{4.47}}
\put(125.00,10.00){\makebox(0,0)[cc]{EA, EB stars}}
\put(43.00,15.00){\circle{0.30}}
\put(56.00,55.00){\circle{0.30}}
\put(57.00,58.00){\circle{0.30}}
\put(63.00,80.00){\circle{0.30}}
\put(79.00,38.00){\circle{0.30}}
\put(47.00,88.00){\circle{0.30}}
\put(116.00,65.00){\circle{0.30}}
\put(43.00,65.00){\circle{0.30}}
\put(42.00,67.00){\circle{0.30}}
\emline{46.00}{116.00}{49}{98.00}{3.00}{50}
\put(76.00,50.00){\circle{0.30}}
\put(77.00,49.00){\circle{0.30}}
\put(75.00,52.00){\circle{0.30}}
\emline{46.00}{116.00}{51}{32.00}{155.00}{52}
\put(49.00,110.00){\circle{0.30}}
\emline{152.00}{44.00}{53}{46.00}{1.00}{54}
\emline{118.00}{61.00}{55}{61.00}{0.00}{56}
\put(88.00,28.00){\circle{0.30}}
\put(90.00,30.00){\circle{0.30}}
\put(85.00,26.00){\circle{0.30}}
\put(99.00,40.00){\circle{0.30}}
\put(101.00,43.00){\circle{0.30}}
\put(98.00,40.00){\circle{0.30}}
\put(97.00,37.00){\circle{0.30}}
\put(78.00,19.00){\circle{0.30}}
\put(78.00,16.00){\circle{0.30}}
\put(94.00,12.00){\circle{0.30}}
\end{picture}
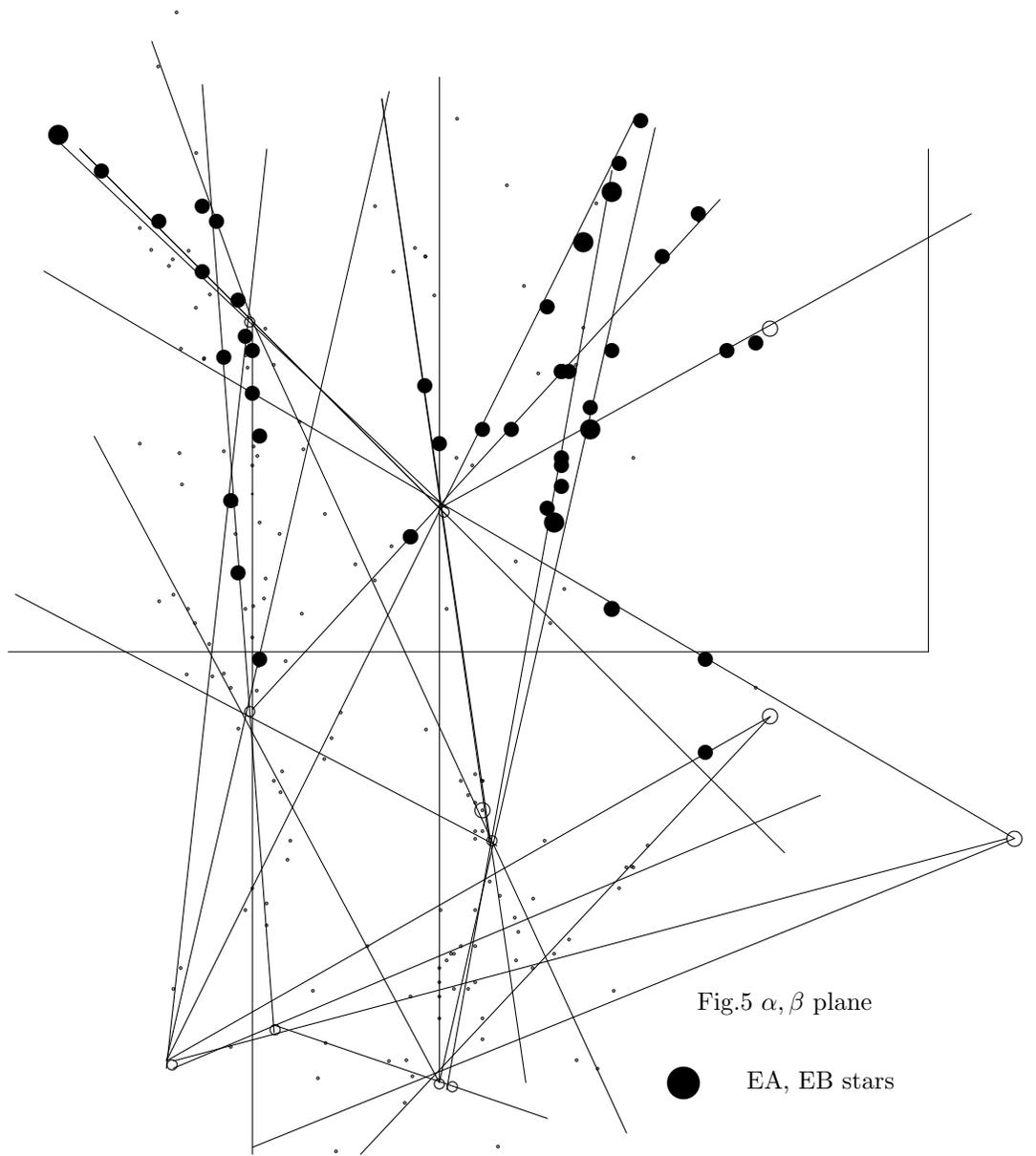
\end{document}